\documentclass[twocolumn,10pt]{article}
\usepackage{usenix}

\usepackage{amsfonts,amsmath,amssymb,amstext,latexsym}
\usepackage{array}
\usepackage{balance}
\usepackage{courier}
\usepackage[inline]{enumitem}
\usepackage{graphicx}
\usepackage{hyperref}
\usepackage{xspace}
\usepackage{listings}

\newcommand*{\cf}{\textit{cf.}\@\xspace}
\newcommand*{\eg}{\textit{e.g.}\@\xspace}
\newcommand*{\ie}{\textit{i.e.}\@\xspace}
\makeatletter
\newcommand*{\etc}{%
    \@ifnextchar{.}%
        {etc}%
        {etc.\@\xspace}%
}
\newcommand*{\etal}{%
    \@ifnextchar{.}%
        {et al}%
        {et al.\@\xspace}%
}
\makeatother

\title{Charting an Intent Driven Network}

\author{
{\rm Yehia Elkhatib\textsuperscript{$\bullet$}, Gareth Tyson\textsuperscript{\textdaggerdbl}, and Geoff Coulson\textsuperscript{$\bullet$}}\\
\textsuperscript{$\bullet$}Lancaster University, United Kingdom\\
\textsuperscript{\textdaggerdbl}Queen Mary, University of London, United Kingdom
}

\begin{document}
\maketitle

\subsection*{Abstract}
The current strong divide between applications and the network control plane is desirable for many reasons; but a downside is that the network is kept in the dark regarding the ultimate purposes and intentions of applications and, as a result, is unable to optimize for these.  An alternative approach, explored in this paper, is for applications to declare to the network their abstract intents and assumptions; \eg ``this is a Tweet'', or ``this application will run within a local domain''. Such an enriched semantic has the potential to enable the network better to fulfill application intent, while also helping optimize network resource usage across applications. We refer to this approach as \emph{intent driven networking} (IDN), and we sketch  an incrementally-deployable design to serve as a stepping stone towards a practical realization of the IDN concept within today's Internet.

\section{Introduction}
\label{sec:intro}

A key principle of the early Internet was the provision of a very simple send/receive interface for applications. 
As the complexity of the Internet has grown, and new capabilities been added (multicast, streaming, mobility, etc.), we have attempted to continue to live with this very simple API. 
Although the approach has served us well so far, clear downsides are emerging.
In particular, applications essentially operate in the dark with respect to the capabilities and functioning of the underlying network, and are therefore obliged to include (complex) logic to handle network related events such as faults, performance fluctuations, service changes (\eg mobility), \etc. 
Similarly, the network operator does not understand the needs of applications beyond the issuance of apparently isolated send and receive calls, thus it is unable to optimize performance for applications or to optimally conserve its own resources. 

At the heart of the matter is the inability of the network to see the underlying \emph{intent} of the application. 
Instead, the network only sees a series of seemingly unrelated micro-transactions (send and receive calls).

One solution might be to extend the network API to give direct access to all the individual 
network elements such as caches, middleboxes, routers, \etc.
But this is clearly problematic for many reasons: application developers would find it too hard to understand, the API would regularly mutate in line with corresponding changes in the technologies, and it would promote unforeseen interactions between per-technology API elements, with unpredictable and probably undesirable consequences.

In this paper, we propose \emph{Intent Driven Networking (IDN)} as an alternative approach. It enables the formulation of an application's ``intents'' as high level statements of its predicted macro-level behaviour, \ie an abstract formulation of what it desires from the network, while remaining agnostic about the underlying means used to satisfy them (protocols, \etc).
For example, one particular intent might be to communicate with a particular group of users (\eg collaborative document editing); another might be to stream a video uninterruptedly while switching between a laptop and a smartphone as well as from an 802.11 network to a cellular one. 
Whereas the current Internet sees such intents as sets of isolated micro-transactions, an intent driven Internet would understand their aims, and therefore be in a position to optimize accordingly.

IDN also allows us to simplify the development of applications by removing the need for them to provide the ``cover all cases'' logic. 
Instead, we feed user requirements straight down to the network thus providing flexibility in how different user application requirements are met without predefined restrictions. 
For instance, it might implicitly ensure the availability of a certain service despite the failure of a remote server; or ensure a certain level of Quality of Experience (QoE) even if the application is not designed to seek alternative potentially better routes; \etc
As a consequence, IDN facilitates more fluid development of end user applications and is conducive to better alignment of the network to application needs.

Another good example arises in the context of mobility support~\cite{Tyson13where}. 
Currently, host mobility is an extremely complicated process, typically managed in the lower layers of the network stack by Mobile IP. In essence, these lower layers attempt to hide the effects of host mobility (\eg changes in IP addresses) from the higher layers (\eg applications) using costly mechanisms such as tunnelling. This measure is necessary as nearly all applications assume stable, globally reachable addressing, as well as consistent connectivity, none of which are valid in the case of mobility. 
These needs, however, often only emerge because the lower layers have no understanding of the \emph{intents} that exist at the higher layers. Imagine that an application, for example, has no need for consistent addressing or, alternatively, only requires access to content within the local domain (\ie no need for global addressing). In such circumstances, the constraints are relaxed and Mobile IP's blanket-style behaviour becomes unnecessary. However, due to the network's inability to see the application's intent, there is no way to decide when such principles should or should not be applied. Consequently, the lowest common denominator must always prevail.

\begin{figure}[!htb]
	\centering
 	\includegraphics[width=0.8\columnwidth, trim=5cm 6cm 3cm 3cm, clip=true]{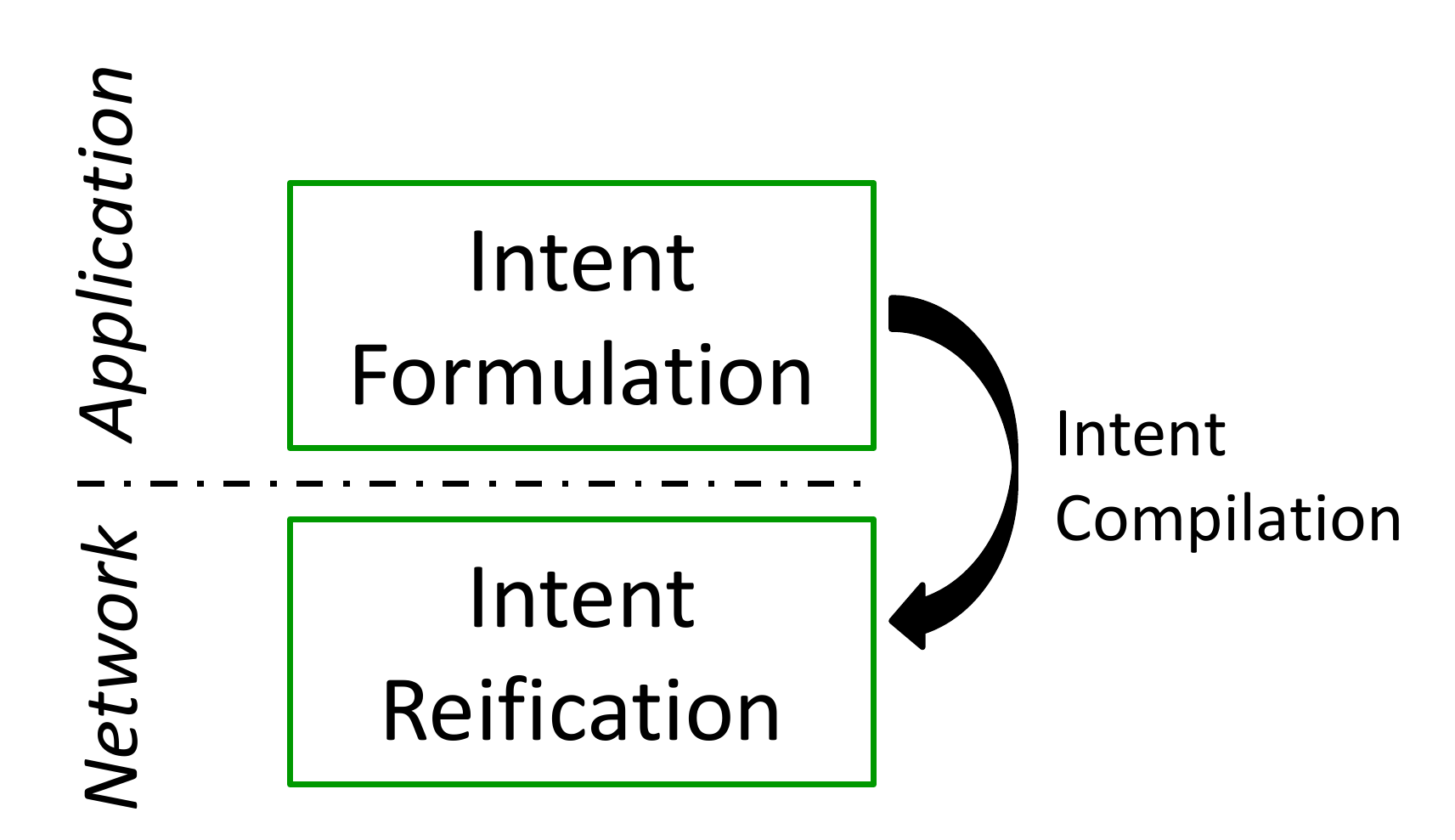}
	\caption{High level view of an intent driven network.}
	\label{fig:arch}
\end{figure}

In order to attain the IDN vision, we need means by which application intents are \emph{formulated}, \emph{compiled}, and ultimately \emph{reified} (\ie acted upon) in the network. The structure implicit in this statement is illustrated in \figurename~\ref{fig:arch}, which serves as the orienting architecture for the remainder of this paper.

We present the concept of an intent driven network using both a straw man design (\S\ref{sec:intents}) and illustrative examples (\S\ref{sec:eg}). We focus particularly on the practical concern of how IDN might be incrementally and partially deployed in the existing Internet without restarting at year zero. We then comment on the implications of this IDN design (\S\ref{sec:impl}) and lay out a roadmap for the incremental realisation of IDN (\S\ref{sec:map}).

\section{Intents}
\label{sec:intents}
This section provides a straw man design of IDN. 
Firstly, we define what an intent is (\S\ref{sec:intents:intent}). 
We then propose an approach for the formulation of intents based on compositions of primitive verbs (\S\ref{sec:intents:ontology}). 
Lastly, we discuss the mechanics of reifying intents in terms of a structure of ``mediators'' in the network (\S\ref{sec:intents:mechanics}).

\subsection{What is an Intent?}
\label{sec:intents:intent}
An \emph{intent} is an abstract declaration of what the application desires from the network on behalf of the user.
It is a composition of a set of primitive ``verbs'', each describing a specific but high-level operation.
For example, an intent to update an Instagram feed might be composed of primitive verbs to reconfigure the application topology (connect to a service and to peers), exchange data (update the content), and uphold a certain QoE level (allocate sufficient network resources).
In response to this, the network carries out the necessary configuration in order to best serve such an intent. This could entail setting up direct connections between users, and allocating fair shares of router queues considering other services in the network.

In more detail, the primitive elements that comprise intents are expressed as \emph{$<$verb, object, modifiers, subject$>$} tuples. 
A \emph{verb} is an operation that describes the intent based on an ontology (described next in \S\ref{sec:intents:ontology}). 
\emph{Object} identifies a service, process or item that is the objective of the verb. 
\emph{Modifiers} are then used to specialize or parameterize this; each modifier can be tagged as either `essential' or `desirable', indicating prioritization preference.
\emph{Subject} is an (optional) identifier of another service/process/item that is to be linked to that defined in \emph{object}. 

Essentially, intent expression is based on the \textit{verb-object-subject} sentence structure used in linguistics, supplemented by \emph{modifiers} as an additional set of words. 
Primitive intents expressed using such sentences are then composed using recursive encapsulation to form a full intent. 

Intents are not limited to only user applications; they extend to applications operated by other players in the network (such as ISPs, cloud service providers, content providers) to express their own intents.

\subsection{Formulating Intents}
\label{sec:intents:ontology}

\begin{figure}[!htb]
	\centering
	\includegraphics[width=\columnwidth, trim=1.5cm 2.8cm 6cm 0.8cm, clip=true]{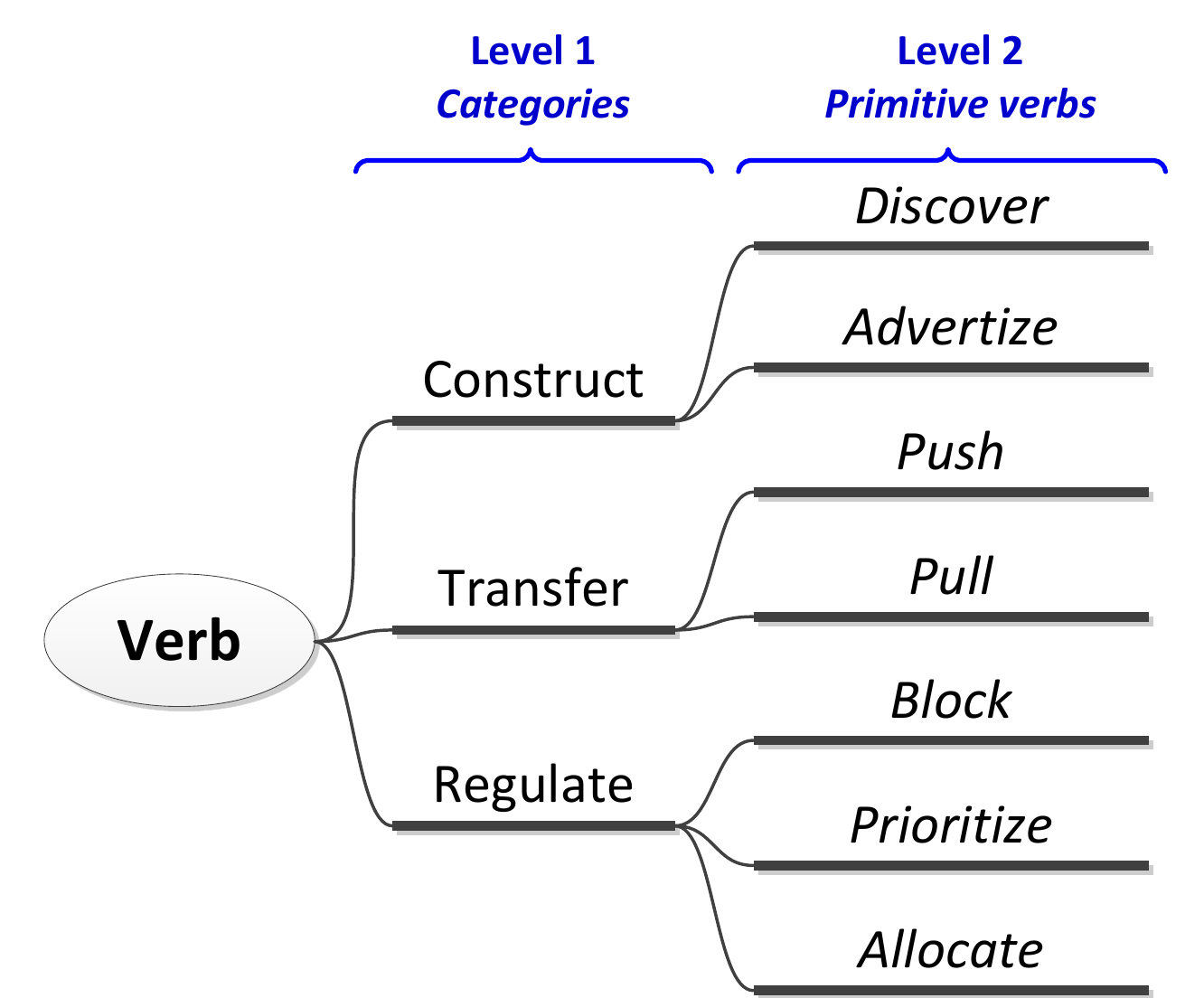}
	\caption{A basic ontology of primitive verbs.}
	\label{fig:dictionary}
\end{figure}

An intent verb is expressed using one of the ontology entries in \figurename~\ref{fig:dictionary}. 
This is not a comprehensive ontology; modification and expansion is possible through collaboration with the wider systems research community.
The ontology is divided into three operation categories: \emph{Construct}, \emph{Transfer}, and \emph{Regulate}. 
Each of these categories has a number of sub-operations from which the verb is chosen. 
Categories are just logical groupings; it is the verbs that signify the primitive intent.

\emph{Construct} is used when an application needs to form connections to another application (the \emph{object}) in a peer-to-peer fashion, either locally over a broadcast address or remotely. An example is a request of intent for a VoIP client to connect to another VoIP client. 
\emph{Discover} is issued to look for certain applications, whilst \emph{Advertize} allows an application to announce a new service that is able to serve the intents of other applications. Examples include nodes spawning a caching or load balancing service. 

\emph{Construct} intents can be extended to another level that defines various applications to be discovered or advertized, such as data storage, caching, compression, \etc  This would enable an application to offload common processing tasks to the network. 
These enable an application to dynamically employ external third party modules without the latter being a component of the native application code. 
This is of particular use to applications running on machines with scarce resources, like a mobile gaming application offloading its transcoding processes.

\begin{table*}[!bt]
	\centering
	\caption{Examples of different ways to refer to content in a \emph{Transfer} intent.}
	\label{tab:content-refs}
	\begin{tabular}{ll}
		 &   \\
		URL & \texttt{http://releases.ubuntu.com/15.04/ubuntu-15.04-server-i386.iso} \\
		CCN & \texttt{ubuntu.com/torrent/ubuntu-15.04-server-i386.iso} \\ 
		BitTorrent & \texttt{05E965AC45FF0D739B3B8998FFFB815D1F238DE9} 
	\end{tabular}
\end{table*}

\emph{Transfer} intents allow applications to pull and push content (the \emph{object}). 
An item of content could be referred to in any of a number of different ways, as illustrated in Table~\ref{tab:content-refs}.
In this sense, a \emph{Transfer} intent is analogous to an ICN abstraction, where the \emph{Push} verb corresponds to a prefix announcement whilst \emph{Pull} corresponds to an interest packet. 

Finally, \emph{Regulate} intents capture the desire of an application to have traffic handled in a certain way in the network rather than locally. 
This is helpful for propagating traffic management higher up the network closer to the source, which facilitates better network management and aggregation of interests.
An example is an intent to block ssh login attempts from a certain address block, or to prioritize traffic from a service like \texttt{hulu.com}, or to apply QoE fairness policy (\eg~\cite{fhmn}).

\subsection{Reifying Intents}
\label{sec:intents:mechanics}

Our conceptual architecture relies on a hierarchical structure of mediators deployed in the network. 
These are middleboxes that arbitrate between user intents, network and service operator policies, and the current state of the network. 
We refer to this mediation presence in the network as \emph{Maat} and each of the middleboxes as a \emph{Maat} agent, in reference to the ancient Egyptian concept of conflict resolution to achieve harmonious equilibrium and order.

Given this, user intents are initially sent on a specific broadcast address to be picked up by a local Maat agent.
If a Maat agent is not available as signified by the expiry of a timer since issuing the intent, the application can widen the address scope to seek another agent in the parent subnetwork, or alternatively it could choose to fall back to non-IDN behaviour.

If a Maat agent finds that it is able to satisfy an intent, its job is to ``reify'' this intent by deploying or activating the required mechanisms (such as an in-network function) or identifying candidate services (a nearby deployment), and consequently sending the relevant information back to the user application to realign itself accordingly. 
The Maat agent is also required to create a session to keep track of how the intent was met. This is important for auditing mediation efficiency.

In realizing IDN we do not propose the re-writing of the entire network stack\footnote{Although an IDN architecture could indeed be approached from this angle, it is likely to create excessive disincentives that limit its deployment.}. Instead, we propose to overlay the concept of intents onto the existing Internet architecture. As such, our straw man IDN architecture has been designed to support \textbf{backward compatibility} (falling back to non-IDN behaviour) and \textbf{incremental/ partial deployment}. We will discuss this further in \S\ref{sec:impl:chl}.

\section{Examples}
\label{sec:eg}
We now provide a set of use case examples to further illustrate the formulation of intent and its corresponding reification. 

\subsection{Use case \#1: Clustering}
Imagine that Alice is editing a document with her colleagues Bob and Charlie on the Google Docs cloud service. Currently, this is handled in a manner that is analogous to a chat server where the collaborators connect to the Google backend to push edits and/or receive updates. 
The problem with this is that it involves unnecessary communication back and forth to where the remote service is hosted when some or all of the collaborators may be within a short network distance of each other.

Using IDN, the local application on Alice's device (Google Docs in this case) would express its intent to share updates between Alice and a set of other users.
For that, it needs to communicate with the Google backend to fetch the addresses of the collaborators' devices. These are then used to formulate an intent as follows: 
\begin{verbatim}
<allocate,
  ip_multicast,
  (ttl=32,essential),
  <discover, 
    GoogleDocs,
    (userID=92cd701c0be,essential),
    (userID=33a88280853,essential),
    NULL
  >
>
\end{verbatim}
Note that the intent here takes the form of a composition between \emph{discover} and \emph{allocate} verbs. Having first discovered the various players in this scenario (the GoogleDocs application and the relevant users), the network allocates a multicast group, and Maat responds with the group address.
From then onwards, local communications are exchanged over this group, and Alice's application is responsible for sending periodic updates to the Google backend for backup if it requires. If the collaborator devices move or change, Alice will issue a new intent accordingly.

\subsection{Use case \#2: Discovery}
Serendipitous peer discovery is important for emerging Internet applications. 
A particularly important application of discovery is in Internet of Things (IoT) environments where a large number of hosts would be operating different services in any one locality. 
Currently, discovery relies on the presence of directory or similar services, which obviously has its limitations in terms of consistency, scalability (considering the scale of IoT swarms to come~\cite{Lee2014swarm}), and crossing vendor and operator divides~\cite{elkhatib2015hotcloud}.

In such a context we might signal an intent to build a new overlay structure from a set of suitable nodes (\eg~\cite{tectons,holons}). This would work in a fashion similar to ARP; a node would seek other nodes that fit certain criteria on the service(s) they operate, location, communication mode, QoS metrics, \etc 
The network would then propagate this intent announcement according to the criteria laid out in the intent modifiers. 

Consider for instance an actuator in an IoT deployment that wants to find a nearby node capable of running a MapReduce analytics workflow over a collection of sensor data. 
In this case, the intent is formed by composing a \emph{discover} primitive verb with a \emph{push} one as follows:
\begin{verbatim}
<push,
  dataset-201507-1800,
  (net=1.2.3.0/24,essential),
  <discover,
    hadoop,
    (rtt<50ms,desirable) &
      (rtt<80ms,essential),
    hadoop-workflow.jar
  >
>
\end{verbatim}

This would be examined and collated by Maat in order to allow the intent to be expanded and traverse across different networks and operational environments (if within the specified criteria). 
As another example, an intent could emanate from a node in a sensor network seeking secure data storage, and use IDN to explore options as diverse as local fixed-power nodes and remote data centers. 
Such discovery may also extend beyond IoT and include intents formed at different levels, such as the ability to choose which middlebox (intrusion detection, proxies, interceptors, anonymizers, \etc) to go through.

It is important to explicitly note that this is \emph{not} a proposal for another discovery broadcast protocol. On the contrary, a key objective of IDN is to enable applications to express high level intents and be agnostic about the protocols used to satisfy them.

\subsection{Use case \#3: Edge Deployment}
Finally, let us consider an example involving different stakeholders: content and service providers.
Both of these stakeholders have a lot to gain from a strong presence towards the edge of the network in areas where there is demand for their services. 

Consider for instance a content provider that finds an increase in the consumption of certain content (say the feature film ``A Beautiful Mind'' following the death of John Nash) in a particular area (say large metropolises in the US). It is in the provider's interest to provide good viewing QoE for its customers and at the same time manage increased load on its backend services. Accordingly, it might decide to push copies of the content to cache in different cities. 

In this case, the intent will be expressed as a composition of a verb that discovers suitable caching services (the \emph{object}) in certain locales (the \emph{modifiers}), a verb that pushes content to the discovered caching points, and a final verb to announce the new content once cached.

The full intent will look like this:
\begin{verbatim}
<push,
  ABeautifulMind,
  (auth=https://provider.com/oauth),
  <push,
    831FD96B0.mp4,
    NULL,
    <discover,
      cache,
      (asn=123456,essential),
      NULL
    >
  >
>
\end{verbatim}
where \texttt{asn} represents the AS number which signifies a certain customer base. 
Other modifiers could be used to identify target locales at a finer grain. 
In a similar fashion, a service provider might deploy applications to nodes offering hosting services. This could be to balance load at the edge, mitigate flash crowds, or improve user QoE.

\section{Implications}
\label{sec:impl}

As already described, IDN pushes some of the meta-logic of a deployed application to the network in a form that can be reified by Maat. 
As such, IDN opens up a whole new set of opportunities in research and operational circles, and also creates some challenges. We now discuss some of these.

\subsection{Opportunities}
\label{sec:impl:opp}
IDN opens up self-adaptation opportunities for all players in the network space, \eg users, developers and service providers. 
Users benefit from improved QoE through service provisioning that is dynamic and adaptive to their requirements and contexts. 
Application developers gain access to higher programming primitives that facilitate fluid application behaviour at runtime, with less reliance on ad hoc means of connecting services and mitigating failures.
Service providers are empowered to provision their services in a migration-ready form so that they will be able to compete to provide better QoE for their end users. 

IDN also opens a market for hosting services towards the edge. This can be beneficial particularly for small and medium sized service providers who cannot afford a highly customized CDN presence like the Googles and Facebooks of the world. Instead, they would be able to bid for edge resource provisioning that in many parts of the world has a wider reach that traditional CDNs~\cite{elkhatib2015building,Fanou2016frontier}. 
As such, there is also room for intermediaries to broker between the above parties.

\subsection{Challenges}
\label{sec:impl:chl}
With the benefits that IDN brings, it also generates a number of challenges. The most prominent of these are \emph{trust} -- specifically in terms of \emph{security} and \emph{efficiency} -- and \emph{deployment}.

The \textbf{security} challenge could be summarized by the following question: \textit{Could the application trust the network to interfere with its communications, potentially redirecting it to an unintended destination?}
This is indeed a major challenge that we recognize. We should first clarify that the in-network Maat agents receive and compile intents, not the subsequent communication which is more likely to contain sensitive information.
Based on this, Maat would have information about the desires of the application such as connecting with peers, advertizing services or content, regulating network traffic, \etc
There is potentially a lot of risk in divulging such information to outside parties.
It is also noteworthy that such challenges are also being faced by the current Internet architecture.

The \textbf{efficiency} concern is summarized with a subtly different question: \textit{Would the application trust the network to potentially impede or interfere with its performance?} 
Maat will have significant influence on where the application is redirected to serve its intent. 
As far as the application is concerned, Maat mediators are black boxes that might have interests conflicting with those of the application users. They could also be misconfigured, resulting in non-optimal mediation.
We perceive this challenge to in fact be an opportunity for \emph{auditing schemes} that ensure the efficiency of mediation. 
For this, we envisage regular reporting of intent, and resulting ``mediation logs'' that could be scrutinized to ascertain efficiency. In a multi-mediator market, the mediation score resulting from such auditing mechanisms would engender competition.

Another challenge relates to \textbf{deployment} and scalability.
The core IDN design lends itself to partial deployment through independent rollout of Maat agents most likely at the edge. There are different ways of doing this, one of which is to augment wireless routers with additional modules. Such devices, however, are typically resource constrained and might suffer from performance issues if a large number of services are advertized on their local address spaces.
One way of avoiding this is to deploy dedicated Maat agents instead of piggybacking them on existing infrastructure. This comes with its own cost, but is fairly feasible using commodity hardware Linux boxes.

\section{Related Work}
\label{sec:rw}

We discuss work that is relevant to our IDN concept, and briefly discuss some of the technologies that will enable the reification of intents in the network.

\subsection{x-Centric Approaches}

Bringing application awareness to networks has been a long sought after goal, with a number of technologies and network architectures being presented. 

\textbf{Resource-centric. }
The Representational State Transfer (REST) architectural principle \cite{fielding2000rest} reduces network interactions to a few \emph{verbs} (GET, POST, DELETE, \etc). 
REST professes an entirely stateless, resource-oriented architectural style, transitioning between states using the data included in the requests. This makes infrastructure scalability and manageability much easier.
However, the REST philosophy continues to adopt the ``narrow'' network API approach and, thus, continues to suffer from all of the associated problems discussed in \S\ref{sec:intro}.

\textbf{Network-centric. }
Information-centric networking (ICN) solutions have been proposed to convert networks into inherent content delivery systems~\cite{Jacobson09}. Similarly, service-centric networking (SCN)\footnote{Also known as service-oriented networking.}~\cite{freedman2010service,Braun2013scn,Griffin2014son,Tschudin2014nfn,Sathiaseelan2015scandex} extends ICN principles to apply to services as well as content. 
Both ICN and SCN attempt to 
break away from statically binding to specific network resources. However, they only partly address the problems we have outlined in the specific cases of accessing content/services: they do not naturally generalize to other scenarios, \eg those involving switching of networks.

\textbf{Stakeholder-centric. }
The Experience-oriented network architecture (EONA)~\cite{Jiang2014eona} is one in which application and content providers as well as infrastructure operators exchange information from their respective control loops to improve user experience. 
We take inspiration from EONA, but are concerned about the viability of its approach.
In a world where data is the new oil, we can not imagine such cooperative exchange of information  happening between parties with sometimes conflicting interests~\cite{clark2014measurement}.
The authors do not provide a reasonable argument for how this would be realized.

\textbf{Application-centric. }
Closer to our proposal are recent efforts in the direction of enabling applications to express their requirements and allowing these to percolate down to the underlying network. 
Pyretic~\cite{reich2013modular} is an open source Python framework that raises the level of abstraction of writing network policies, enabling the definition of sophisticated network structures through a high-level language. 
Merlin~\cite{Soule2013Merlin} is another declarative language that enables the specification of a global networking policy, which is expressed as a collection of logical predicates to identify traffic subsets and a set of statements indicating the action(s) to be taken on each subset. 
However, both Pyretic and Merlin focus on issues relating to unifying network administration rather than identifying and addressing application requirements.

Other relevant efforts include: 
yanc~\cite{Monaco2013yanc}, an abstraction in Linux to facilitate network control logic to be written in any language;
FlowOS~\cite{Bezahaf2013flowos}, a programming model to capture and process Internet flows;
NOSIX~\cite{Yu2014nosix}, an abstraction layer to enable portable deployments; and 
P4~\cite{Bosshart2014p4}, a language to configure switches to process packets and match header fields.

\subsection{Enabling Technologies}

The \emph{Active networking} (AN) paradigm enables users to modify network behavior by sending custom code to be executed on network devices (\cf \cite{Tennenhouse1997survey}). 
\emph{Software defined networking} (SDN) removes the need for bit-wise configuration of network components, and instead allows a central policy to be applied system-wide.  
Both AN and SDN technologies could be used to the benefit of users, 
but neither helps the application signify its needs. 
Thus, they are complementary to IDN by providing mechanisms to manage and modify the network in order to reify intents.

\section{Summary and Roadmap}
\label{sec:map}

\begin{figure*}[!ht]
	\centering
	\includegraphics[width=0.86\textwidth, trim=0 1.3cm 0 8.4cm, clip=true]{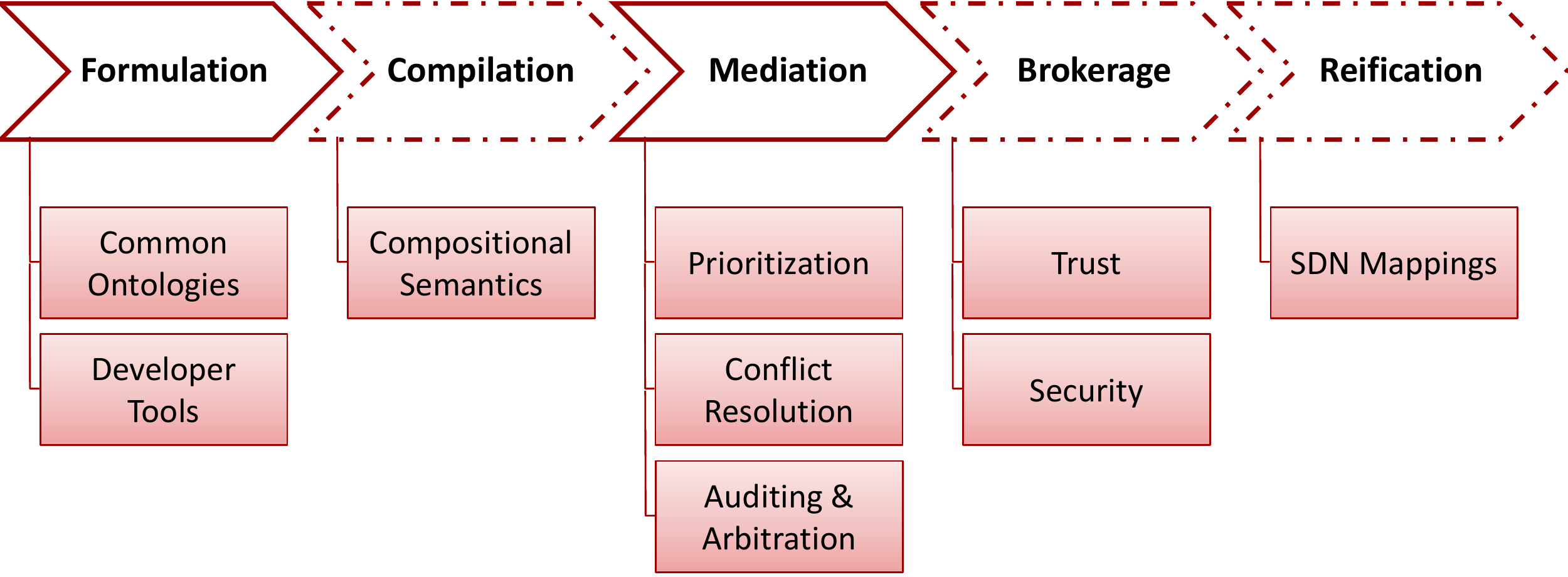}
	\caption{A roadmap for the realization of IDN.}
	\label{fig:roadmap}
\end{figure*}

We have proposed the concept of Intent Driven Networking (IDN) in which applications and also other players such as content providers formulate their communication-related ``intents'' in high-level terms that get transformed into network-level reifications that better support the application's declared intents. 
We put forward a straw man design which specifies how  intents might be formulated (\S\ref{sec:intents:intent}). Formulation involves an ontology of intent verbs to signify various application desires (\S\ref{sec:intents:ontology}). Reification relies on the Maat system that provides in-network mediation between user intents and policies of network and service operators (\S\ref{sec:intents:mechanics}).

Apart from enjoying network service levels that better match their intents, applications also benefit from IDN in that some of their logic could be pushed to the network. No longer are they expected to ship with intricate conditional logic to work around unexpected network behaviour. (They still could employ such logic, but they would thereby be limiting their ability to be deployed in foreign environments and under unforeseen conditions.) 

Maat also exploits the high semantic content of intent specification to optimize resource usage within the network. It does this by explicitly negotiating how each intent might best be reified to produce the most ``friendly'' way in which each network interaction can be performed, considering the needs of all stakeholders.

There is a huge body of future work required to develop IDN into a viable implementation. 
We tried to capture some of the required next steps in a roadmap depicted in \figurename~\ref{fig:roadmap}. This, however, is undoubtedly a non-exhaustive plan of action.
Therefore, we solicit contributions from the wider systems research community, architects and developers of different disciplines.

We see the process of developing a proof-of-concept realisation of our design as falling into three main workpackages. First, we will work on the formalism of intent specification and its compilation into a format that can be used by Maat. This will involve refining the ontology presented in this paper and defining a domain-specific language for the formulation of intents on the basis of this, and also the creation of associated developer tools. It will also involve investigating the semantics of the (recursive) composition of intents in terms of existing intents and ultimately in terms of primitive verbs. Composition semantics must take into account the effects of composing network functions that interact in complex ways: for example how caching strategies change when associated with mobility. 

The second workpackage involves the definition of the negotiation protocol employed between Maat agents. This should be independent of any particular set of verbs and rely on generic notions of utility and priority as derived from intent specification, and be capable of handling negotiations between multiple stakeholders and converging on distributed consensus. It is clear that mediation is a highly complex task as it is likely that many conflicts will emerge. For example, a user streaming content would want high quality delivery at low cost, a publisher would wish to have their content viewed as many times as possible, and an ISP would prefer to only have low-cost (locally available) content viewed. Such potentially conflicting viewpoints will need to ensure thorough negotiation to ensure that all stakeholders are incentivized to cooperate in the scheme.

The third workpackage would then focus on practical aspects of the deployment of Maat agents, as discussed in \S\ref{sec:impl:chl}.

As a final consideration, marketplace brokerage is an area with a lot of potential for reifying spontaneous and strategic intent.
Reification is likely to create the need for running in-network services towards the edge. 
Marketplaces of resources to host such services might benefit from the operation of brokerage and arbitrage agencies, a role which might be co-located with Maat. 
For this, thorough investigation is required to alleviate concerns regarding trust and security.
Efforts are also sought for reifying mediation outcomes in the form of adjusting the network control plane (using SDN technologies) or providing information that could be used 
for late-binding. 

We hope that our proposed architecture will serve as an initial step towards a long-term research campaign that focuses on the higher-layer wants and needs of Internet stakeholders rather than forcing them into using fixed and constrained abstractions.

\balance{
	\bibliographystyle{acm} 
	\bibliography{intent}
}

\end{document}